\documentstyle[sprocl]{article}
\def\Journal#1#2#3#4{{#1} {\bf #2}, #3 (#4)}

\def\NPB{{\em Nucl. Phys.} B}
\def\PLB{{\em Phys. Lett.}  B}
\def\PRL{\em Phys. Rev. Lett.}
\def\PR{{\em Phys. Rev.}}
\def\PRD{{\em Phys. Rev.} D}
\def\PRB{{\em Phys. Rev.} B}
\def\ZPC{{\em Z. Phys.} C}
\def\JETPL{{\em JETP Lett.}}
\def\CMP{{\em Commun. Math. Phys.}}

\newcommand{\be}{\begin{equation}}
\newcommand{\ee}{\end{equation}}
\newcommand{\bea}{\begin{eqnarray}}
\newcommand{\eea}{\end{eqnarray}}

\newcommand{\cK} {{\cal K}}
\newcommand{\mod}{({\rm mod}2)}
\newcommand{\hf} {{1\over2}}

\newcommand{\nonu}{\nonumber\\}

\newcommand{\adots} {{\mathinner{\mkern2mu\raise1pt\hbox{.}\mkern2mu
\raise4pt\hbox{.}\mkern2mu\raise7pt\hbox{.}\mkern1mu}}}

\newcommand{\plaq}{\setlength{\unitlength}{.5cm}\raisebox{-.2cm}{
   \begin{picture}(1.2,1.2)(-.6,-.6)
   \basispl\basisar
   \put(-.5,-.5){\circle*{.2}}
   \put(-.55,-.55){\makebox(0,0)[tr]{\footnotesize $x$}}
   \put(-.55,0){\makebox(0,0)[r]{\footnotesize $\nu$}}
   \put(0,-.55){\makebox(0,0)[t]{\footnotesize $\mu$}}
   \end{picture}}}

\newcommand{\basispl}{
   \put(-.5,-.5){\line(1,0){1}}
   \put(.5,-.5){\line(0,1){1}}
   \put(.5,.5){\line(-1,0){1}}
   \put(-.5,.5){\line(0,-1){1}} }

\newcommand{\basisar}{
   \put(0,-.5){\vector(1,0){0}}
   \put(.5,0){\vector(0,1){0}}
   \put(0,.5){\vector(-1,0){0}}
   \put(-.5,0){\vector(0,-1){0}} }

\newcommand{\twooneplaq}{\setlength{\unitlength}{.5cm}
   \raisebox{-.2cm}{
   \begin{picture}(2.2,1.2)(-1.1,-.6)
   \put(-1,-.5){\line(1,0){2}}
   \put(-1,.5){\line(1,0){2}}
   \put(-1,-.5){\line(0,1){1}}
   \put(1,-.5){\line(0,1){1}}
   \put(-0.5,-.5){\vector(1,0){0}}
   \put(0.5,-.5){\vector(1,0){0}}
   \put(-0.5,.5){\vector(-1,0){0}}
   \put(0.5,.5){\vector(-1,0){0}}
   \put(-1,.0){\vector(0,-1){0}}
   \put(1,.0){\vector(0,1){0}}
   \multiput(-1,-.5)(1,0){3}{\circle*{.2}}
   \multiput(-1,.5)(1,0){3}{\circle*{.2}}
   \put(-1.0,-.55){\makebox(0,0)[tr]{\footnotesize $x$}}
   \put(-1.0,0){\makebox(0,0)[r]{\footnotesize $\nu$}}
   \put(0,-.55){\makebox(0,0)[t]{\footnotesize $\mu$}}
   \end{picture}}}

\newcommand{\twoplaq}{\setlength{\unitlength}{1cm}\raisebox{-.5cm}{
   \begin{picture}(1.2,1.2)(-.6,-.6)
   \basispl
   \put(-.5,-.5){\circle*{.1}}
   \put(-.5,.5){\circle*{.1}}
   \put(.5,-.5){\circle*{.1}}
   \put(.5,.5){\circle*{.1}}
   \put(0,-.5){\circle*{.1}}
   \put(0,.5){\circle*{.1}}
   \put(.5,0){\circle*{.1}}
   \put(-.5,0){\circle*{.1}}
   \put(-.25,-.5){\vector(1,0){0}}
   \put(.25,-.5){\vector(1,0){0}}
   \put(.5,-.25){\vector(0,1){0}}
   \put(.5,.25){\vector(0,1){0}}
   \put(-.25,.5){\vector(-1,0){0}}
   \put(.25,.5){\vector(-1,0){0}}
   \put(-.5,-.25){\vector(0,-1){0}}
   \put(-.5,.25){\vector(0,-1){0}}
   \put(-.55,-.55){\makebox(0,0)[tr]{\footnotesize $x$}}
   \put(-.55,0){\makebox(0,0)[r]{\footnotesize $\nu$}}
   \put(0,-.55){\makebox(0,0)[t]{\footnotesize $\mu$}}
   \end{picture}}}

\begin{document}
\title{THE ANTI-FERROMAGNETIC VACUUM}
\author{JANOS POLONYI\footnote{polonyi@fresnel.u-strasbg.fr}}
\address{
Laboratory of Theoretical Physics, Louis Pasteur University\\
3 rue de l'Universit\'e 67087 Strasbourg, Cedex, France\\
\vspace*{.5cm}
Department of Atomic Physics, L. E\"otv\"os University\\
Puskin u. 5-7 1088 Budapest, Hungary}
\maketitle\abstracts{Certain effective vertices may
generate a non-homogeneous, periodic vacuum structure. The excitations above 
such a vacuum are studied in the framework of the $\phi^4$ and 
gauge models. The formation of the non-homogeneous vacuum is
accompanied by the dynamical breakdown of the space-time inversion
symmetry. Chiral transformation is introduced for bosons in close 
analogy with lattice fermions.}
\section{Introduction}
The ferromagnetic condensate is a well known device for mass
generation in particle physics ever since the seminal work of
Nambu and Jona-Lasinio\cite{najo}. This condensate
is a coherent state of particles with vanishing momentum. We suggest
in this talk the use of another type of condensate in Quantum Field Theory
which is a coherent state consisting of particles with non-zero momentum. The
field operator which has non-homogeneous vacuum expectation value in this
vacuum oscillates with the characteristic momentum of the
particles of the condensate. Based on the formal similarity with the Ne\`el state 
of the anti-ferromagnetic Ising model this condensate will be called 
anti-ferromagnetic. 

The realistic theories of Particle Physics are effective models since we
do not know the physics up to infinite energies. Each effective theory
has an energy range where it is supposed to be applicable and contains
effective interactions due to the particle exchange processes
occuring beyond its energy regime. The impact of these effective
vertices is usually classified by the Appelquist-Carazzone decoupling
theorem\cite{apca}, which states that the non-renormalizable effective
vertices are suppressed by the ratio of the light and the heavy particle
mass. This result is usually taken as an indication that the effect of 
these vertices whose number is uncontrollably large is numerically small
and it is enough to take into account the contributions of the heavy particle
exchange processes to the renormalizable vertices only as far as the low
energy physics is concerned.

It is overlooked in this argument that the decoupling theorem is
based on a careful analysis of the loop corrections and one tacitly assumes
that nothing important is going on at the tree level. Can it happen that the
tree level semi-classical vacuum is modified by the heavy particle
exchange contributions ? The dependence of the solution of the classical 
equation of motion on the non-renormalizable coupling constants is more complicated 
than those of the radiative corrections. The strategy for the 
classification of the impact of the effective vertices which is based on the decoupling 
theorem is not adequate to estimate the tree level effects of the heavy
particle exchange processes. 

Let us consider the strong and the 
electromagnetic interactions at finite barion number density as an example.
The heavy and light matter particles of this system are the nuclei and the
electrons, respectively. For the low energy phenomena it is enough to consider
the electrons and the slightly ionized atoms as matter particles. Imagine the 
effective theory for the electrons and the photons after the perturbative elimination 
of the ions. We know very well that the vacuum of 
this effective theory is the non-homogeneous solid state lattice, more precisely
its electric field and the corresponding Dirac sea for certain barion densities. 
This non-trivial vacuum can not be generated by the
radiative corrections of the non-renormalizable effective vertices which are strongly 
suppressed according to the decoupling theorem. Instead, it is believed to
be a semi-classical, tree level effect\cite{wigner}, which is happened to be much 
more important than one would have expected according to the order of magnitude of 
the non-renormalizable effective vertices. 

One of the usual strategies to "derive" the solid state lattice is based on the 
Born-Oppenheimer approximation where one eliminates the light electrons and 
describes the semi-classical motion of the heavy ions in the resulting
non-local potential. It is more promising to eliminate the heavy degrees of 
freedom first and study the dynamics of the light particles in the presence of the
effective local interactions later. We believe that the perturbative elimination of the
ions will give us the effective theory whose semi-classical vacuum is the
electric field of the solid state lattice.

This talk is devoted to some general remarks about the appearance of the
non-homogeneous, anti-ferromagnetic vacuum in some models. We
start with the simplest case, the $\phi^4$ model in Section 2. 
The interpretation of the tree level
phase transition is the subject of Section 3. The particle content 
is discussed in Section 4 by paying special attention to the space-time 
inversions. The generalization of the anti-ferromagnetic vacuum
for gauge theories and some of its characteristic phenomena are discussed in 
Section 5. Finally Section 6 is for the conclusion. 

\section{Tree-level non-decoupling for the $\phi^4$ model}
Consider the Euclidean effective action
\be
S[\phi(x)]=\int d^dx\biggl(\hf\partial_\mu\phi(x)
\cK(\partial^2)\partial_\mu\phi(x)+V(\phi(x))\biggr)\label{phac}
\ee
for the scalar field $\phi(x)$ where the potential energy is symmetrical, 
$V(\phi)=V(-\phi)$. In order to arrive at a renormalizable theory we choose
$V(\phi)={m^2\over2}\phi^2+{g\over4!}\phi^4$. 
The higher order, apparently non-renormalizable terms of the kinetic energy,
\be
\cK(z)=1+c_2{z\over\Lambda^2}+c_4{z^2\over\Lambda^4},\label{kinen}
\ee
are supposed to arise from some heavy particle exchange at the energy scale 
$\Lambda$. This energy scale plays the role of a smooth cut-off since the
$O(p^{-6})$ propagator cuts off the loop integrals at $p=O(\Lambda)$. The 
coupling constants $c_j$ are irrelevant as far as the quantum fluctuations 
around a homogeneous vacuum are concerned because they always appear
with the suppressing factor $O((p/\Lambda)^j)$ in the radiative corrections.

In order to estimate the tree level effects of $c_j$ we first suppose that
the vacuum remains homogeneous, $<\phi(x)>={\rm const}.$ The quantum
fluctuations, the eigenfunctions of the second functional derivative of the
action are plane waves with the eigenvalue
\be
\lambda(p^2)=m^2+p^2-c_2{p^4\over\Lambda^2}+c_4{p^6\over\Lambda^4}.
\label{invp}
\ee
We shall consider $m^2<0$ in this talk, where one has a ferromagnetic 
condensate, $<\phi(x)>={\rm const}$, for $c_j=0$.
But observe that for sufficiently large $c_2$ the eigenvalue becomes negative
for non-vanishing momentum indicating an instability for particles with
momentum $p=O(\Lambda)$. The coherent state of these particles generates
a field expectation value which oscillates at the characteristic length
$O(\Lambda^{-1})$. The numerical solution\cite{herve} of the Euler-Lagrange
equation of (\ref{phac}) supports this picture and results an
oscillating classical field which is a pure cosinus up to 99\% in $d=1$ and 
a somewhat lower degree in higher dimensions. Such a small spread in the
Fourier space indicates weak interaction between the particles which form
the condensate.

For $|m^2|<<\Lambda$, $3c_4<c_2^2$, $\lambda(p^2)$ takes negative values
and we have anti-ferromagnetic
condensate. The ferromagnetic condensate appears when $3c_4<c_2^2$ 
and $m^2<0$. We found no phase where both $p=0$ and $p\not=0$
condensates are present at the semiclassical level. The situation becomes
more involved in lattice regularization. This is because there are in fact
two lattices in this case, one of those of the anti-ferromagnetism and 
the other is the regulator. One finds several regular phases 
which are labeled by the period length of the anti-ferromagnetic lattice
measured in the units of the cut-off. These phases are separated by
the incommesurable phases where the saddle point shows chaotic
behavior.

\section{Phase structure}
The elementary excitations above the anti-ferromagnetic vacuum 
are the Bloch waves. The free inverse propagator (\ref{invp}) may
have two minima along each axes in the momentum space
so it is natural to introduce $2^d$ regions in the Brillouin zone 
$|p_\mu|\le\pi/a$ around each saddle points of $\lambda(p^2)$,
$p_\mu=P^\alpha_\mu(c_2,c_4)$, $\alpha=1,\cdots,2^d$,  
which together cover the whole momentum space.
If a saddle point is a local minimum then the excitations around it
are particle like. 

It is instructive to consider the theory in lattice regularization. The free 
inverse propagator is
\be
G_0^{-1}(p)=a^2m^2+{\cal P}^2(p)-c_2{a^2\over\pi^2}({\cal P}^2(p))^2+
c_4{a^4\over\pi^4}({\cal P}^2(p))^3,
\ee
where ${\cal P}^2(p)={4\over a^2}\sum\limits_\mu\sin^2{ap_\mu\over2}$.
The key observation is that it is invariant under the transformation 
\be
p_\mu\longrightarrow p_\mu+P^{2^d}_\mu,\label{boch}
\ee
when 
\be
c_2={1\over4d},~~~c_4=0.\label{chpo}
\ee
The vector $P^\alpha_\mu$ is defined by 
$P^\alpha_\mu=P^\alpha_\mu\biggl({1\over4d},0\biggr)={\pi\over a}n_\mu(\alpha)$,
where the components of the vector $n_\mu(\alpha)$ is 0 or 1,
\be
\alpha=1+\sum\limits_{\mu=1}^{d}n_\mu(\alpha)2^{\mu-1}.
\ee
Notice that the lattice action with $c_4=0$ is bounded from below due to 
the $\phi^4$ term in the potential energy \cite{rivasseau}. 

The symmetry of the free propagator reflects the invariance of the kinetic energy
with respect to (\ref{boch}). This transformation introduces a fluctuating sign
for the field variable, $\phi(x)\to\pi\phi(x)$, where the operator $\pi$ acts in the 
space of lattice field configurations as
\be
\pi\phi(x)=(-1)^{\sum\limits_{\mu=1}^d x^\mu}\phi(x)\label{pitr}
\ee
and (\ref{boch}) leaves the potential energy invariant, too. We shall explore the 
consequence of this discrete symmetry of the theory (\ref{chpo}) in two, 
slightly different manners.

The projection operator corresponding to the two sub-lattices of the anti-ferromagnetic
order is $P_\pm={1\over2}(1\pm\pi)$. For the symmetrical theory (\ref{chpo}) we have 
$[\cK(\partial^2)\partial^2,\pi]=0$. Another important property we need is $\pi^2=1$.
These two relations can be used to prove
\be
P_-\cK(\partial^2)\partial^2P_+=0,\label{kede}
\ee
which shows that the field variables of the two sublattices are not coupled
by the kinetic energy. Since the potential energy is ultralocal the two sublattices
decouple and the theory (\ref{chpo}) consists of two identical and independent 
scalar field theories defined on each sublattice.

Suppose that $m^2$ and $g$ are chosen in such a manner that these two theories
are in the symmetry broken phase. Then both of the decoupled fields develop
non-vanishing vacuum expectation values which agree 
in their absolute magnitude and their sign is chosen randomly by a dynamical
symmetry breaking mechanism. In the case when the two condensates
have same (opposite) sign we find ferromagnetic (anti-ferromagnetic)
vacuum in the space-time. 

Let us consider the theory in the vicinity of the symmetric case\cite{devega},
\be
c_2={1\over4d}+a^2M^2,~~~c_4=0.
\ee
The theory is in the ferromagnetic or the anti-ferromagnetic vacuum for
$M^2<0$ or $M^2>0$, respectively. The one-loop analysis\cite{herve} shows 
that the ultraviolet divergences can be removed form the Green functions
by the appropriate fine tuning of the bare parameters, $m^2(a)$, $g(a)$.
The resulting renormalized theory contains two particles
and its vacuum appears homogeneous for observations made at finite 
energies. The two particle species are non-degenerate for $M^2>0$ and as 
we shall show below the space-time inversion symmetry is dynamically broken.
The cut-off independent finite energy physics belongs to
different universality class than the usual $\phi^4$ theory with $c_j=0$.
It remains to be seen if the triviality of the theory persists in this
scaling regime.

The length scale of the condensate could be kept finite as in Solid
State Physics. In this case the hard processes which tend to destroy
the condensate reveal the presence of unitarity violating processes in the theory at
high energies \cite{unit}. This is not a serious problem if the theory
is used only to describe the effects of the low energy excitations. It is
worthwhile noting that the classical solutions may contain ultraviolet Fourier
modes, it is the excitation above the semi-classical, non-fluctuating vacuum what 
is supposed to be at low energy in order the effective theory make sense. We have,
for example, no doubt about the correctness of the description of the low energy
elementary excitations in solids in terms of phonons and electrons. The
hard photon exchanges which may induce lattice defects appear to
violate the unitarity of the phonon-electron system and require the use
of the complete, more fundamental QED treatment. This high energy
violation of the unitarity disappears from the theory where the length scale
of the condensate is renormalized to zero and we expect to find
an acceptable theory without extra anomalies after the renormalization 
around the symmetric point\cite{herve}, (\ref{chpo}).

\section{Chiral bosons}
A more systematical way of exploring the consequence of the symmetry 
(\ref{boch}) can be achieved by exploiting the formal similarities between 
our system and the lattice fermions. To this end 
we introduce the hyper-cube variables $x^\mu=2y^\mu+n^\mu$ and the
corresponding field\cite{morel} $\phi_n(y)=\phi_\alpha(y)=\phi(2y+n(\alpha))$
where $n$ and $\alpha$ will be called the helicity index. 
It is advantageous to introduce the linear superpositions \cite{thun}
$\tilde\phi_\alpha(y)=A_{\alpha\beta}\phi_\beta(y)$, where the matrix
\be
A_{\alpha\beta}=2^{-d/2}(-1)^{n(\alpha)\cdot n(\beta)}
\ee
represents the Fourier transformation on the helicity index.
The Fourier transform of the field $\tilde\phi_\alpha(y)$ in $y$ is 
non-vanishing for $|p_\mu|\le\pi/2a$ and corresponds to the excitations in 
the restricted Brillouin zone around $p\approx P^\alpha$. 
Since 
and 
\be
(A^2)_{n,n'}=2^{-d}\sum\limits_m(-1)^{m\cdot(n+n')}=\delta_{n,n'},
\ee
the inverse transformation is
$\phi_\alpha(y)=A_{\alpha\beta}\tilde\phi_\beta(y)$.

The transformation (\ref{boch}) acts as 
$\tilde\phi_\alpha(y)\longrightarrow\tilde\phi_{2^d+1-\alpha}(y)$
on the Fourier transformed fields and is
represented by $n\to E(0)-n\mod$ in the vector notation. The vector 
$E(k)$, $0\le k\le d$ is defined here as
\be
E_\mu(k)=\cases{1&$k\le\mu$,\cr0&otherwise.}
\ee 

One finds that the propagator has a saddle point in the restricted Brillouin 
zones of $\alpha=2,\cdots,2^d-1$ so the fields $\tilde\phi_\alpha(p)$ with 
these chiral indices do not correspond to particle like excitations\cite{herve}. 
On the contrary, the sectors $\alpha=1$ and $2^d$ describe particle 
like excitations in the vicinity of (\ref{chpo}) and the fields $\tilde\phi_1(y)$ 
and $\tilde\phi_{2^d}(y)$ 
which are mapped into each other by the symmetry transformation are the 
ferromagnetic and the staggered anti-ferromagnetic order parameters, respectively. 

In order to demonstrate the geometrical origin of our discrete symmetry we 
discuss the inversions in space-time. 
The elementary cell of the anti-ferromagnetic 
system consists of the hypercube of $2^d$ points. The inversion 
$I_\mu$ of the coordinate $x^\mu$ flips the $\mu$-th components of the
chiral vector $n_\mu$, 
\be
I_\mu:~~\phi(y)\longrightarrow U_\mu\phi(I_\mu y).
\ee
The matrix $U_\mu$ defined in this manner acts on the helicity index, 
\be
(U_\mu)_{n,m}=\cases{1&if $n_\nu+m_\nu=\delta_{\mu,\nu}\mod$,\cr
0&otherwise.}
\ee
The simultaneous inversion of all space coordinates, 
$P=\prod\limits_{\ell=2}^dI_\ell$, induces the transformation
\be
P:~~\phi(y)\longrightarrow U_P\phi(Py),
\ee
where 
\be
(U_P)_{n,m}=(\prod\limits_{\ell=2}^d U_\ell)_{n,m}=\cases{1&if $n+m
=E(1)\mod$,\cr0&otherwise.}
\ee
By analogy with the spin-half particles the field $U_P\phi(y)$ will 
be called the P-helicity partner of $\phi(y)$. 
The combined effect of the time inversion $T=I_1$ and the space inversion
is represented by
\be
PT:~~\phi(y)\longrightarrow U_{PT}\phi(PTy),
\ee
\be
(U_{PT})_{n,m}=(\prod\limits_\mu U_\mu)_{n,m}=\cases{1&if $n+m
=E(0)\mod$,\cr0&otherwise.},
\ee
and $U_{PT}\phi(y)$ will be called the PT-helicity partner of $\phi(y)$. 

To make the analogy with the fermionic case more complete we shall
determine the transformation rules for $\tilde\phi$ under space-time inversions.
$\tilde U_\mu$, the transformation matrix in the helicity space which
represents $I_\mu$ is given by $\tilde U_\mu A=AU_\mu$, what yields
$\tilde U_\mu=AU_\mu A$. By the help of
\be
(\tilde U_P)_{n,n'}=(AU_PA)_{n,n'}\nonu
=2^{-d}\sum\limits_m(-1)^{m\cdot(n+n')+n\cdot E(1)}
\ee
we find 
\be
(\tilde U_P)_{n,n'}=\delta_{n,n'}(-1)^{n\cdot E(1)}
=\delta_{n,n'}(-1)^{\sum\limits_{\ell=2}^d n_\ell}.
\ee
In a similar manner we have
\be
(\tilde U_{PT})_{n,n'}=\delta_{n,n'}(-1)^{n\cdot E(0)}
=\delta_{n,n'}(-1)^{\sum\limits_{\mu=1}^d n_\mu}.
\ee
The formal similarity of this expression with (\ref{pitr}) reveals that the 
space-time inversion, $PT$, is represented in the helicity space by 
the Fourier transform of our discrete symmetry (\ref{boch}). 
The distinct feature of the theory (\ref{chpo}) is that the P- or the PT-helicity
partners decouple for even or odd values of d, respectively. Such a theory
will be called chiral invariant and its symmetry, (\ref{boch}), is a discrete
version of a boose-chiral transformation. The continuous boose-chiral 
transformations can only be found in scalar field theories with continuous
internal symmetry.
The space-time inversions are diagonal on the chiral field basis of $\tilde\phi$.
The charge with respect to the exchange of the P-helicity partners, the
P-chirality of $\tilde\phi_n$, is defined as 
\be
\chi_P(n)=(-1)^{\sum_\ell n_\ell}.
\ee 
The chiral transformation (\ref{boch}) which differs from that of the fermionic 
models by the presence of the time inversion is diagonal as well, its eigenvalue, the 
PT-chirality, is 
\be
\chi_{PT}(n)=(-1)^{n_0}\chi_P(n).
\ee 

It is interesting to compare this situation with lattice fermions. 
The fermionic theory has $2^d$ non-interacting particle modes whose chiral 
charges sum up to zero. The different helicity partners decouple for the
chiral invariant theory. 
The theory given by (\ref{phac}) displays a species doubling in each 
space-time direction. We introduced $2^d$ fields, $\phi_\alpha(y)$, which 
can be groupped into $2^{d-1}$ "bi-scalars", 
$\psi_\alpha^{PT\pm}(y)=\psi_\alpha(y)\pm U_{PT}\phi_\alpha(PTy)$, or
$\psi_\alpha^{P\pm}(y)=\psi_\alpha(y)\pm U_P\phi_\alpha(Py)$.
These pairs are classified by the help of an $O(2^{d-1})$ flavor group which
acts on the helicity space. The members of the pairs are
exchanged by the discrete transformation P or PT whose analogues are
the Dirac matrices $U_P=i\gamma_0$ or $U_{PT}=\gamma_0\gamma_1\gamma_3$, 
respectively. The particle modes, $\tilde\phi_\alpha(p)$, 
which are the result of the diagonalization of the kinetic energy are the 
eigenvectors of the space-time inversions.
The total chirality charge vanishes, 
$\sum\limits_n\chi_P(n)=\sum\limits_n\chi_{PT}(n)=0$.
One may add further terms to the action in such a manner that the condensate
occurs at $n=E(1)$. This is the Euclidean analogue of the solid state lattice where the vacuum
is periodic in space but constant in time. Such a vacuum breaks the $O(d)$
Euclidean invariance and introduces a $d$-fold degeneracy in the 
spectrum. There is no
species doubling in the time direction and $\chi_P(n)=\chi_{PT}(n)$
in this phase.

We note finally that the decoupling of the P-helicity members at (\ref{chpo}) 
can be see by 
repeating the argument used in the derivation of (\ref{kede}) but replacing the 
operator $\pi$ by $\tilde U_{PT}$. The chiral invariance of the propagator implies
$[\cK(\partial^2)\partial_\mu^2,U_{PT}]=0$. This relation together with
$U_{PT}^2=1$ can be used to prove that the kinetic energy does not
couple the chiral fields with opposite PT-chirality. The potential energy
depends on $\phi_n^2(y)=\tilde\phi_m(y)A_{m,m'}^{(n)}\tilde\phi_{m'}(y)$
where
\be
A_{m,m'}^{(n)}=(-1)^{n\cdot(m+m')}.
\ee
Due to $[A^{(n)},U_{PT}]=0$ any function of  $\phi_n^2(y)$ leaves the opposite
P-helicity partners decoupled.

\section{Gauge theories}
The phase structure of the simple SU(2) model with the extended action \cite{extact}
\be
S=\sum_{x,\mu\ne\nu}\biggl(c_1\plaq+c_2\twooneplaq+c_3\twoplaq\biggr)
\label{exta}
\ee
is remarkable complex \cite{jochen}. This model has a number of phases
which are characterized by different number of anti-ferromagnetic
planes in which the plaquette shows staggered short range order. The fate of
some characteristic features of QCD, such as asymptotic freedom, 
confinement, chiral symmetry breaking is unclear in the new phases. 
It was conjectured
that the phase transitions can be used to construct a continuum theory
with different universality class than for $c_2=c_3=0$. 

The anti-ferromagnetic long range order is characteristic of the semi-classical
approximation. The quantum fluctuations may destroy the long range correlation
leaving only a staggered short range order behind. This is supposed to
happen if the theory is infrared unstable as in the case of asymptotically
free models or spin liquids \cite{fradkin}. The vacuum recovers the
usual space-time translation symmetries after the long range order
is removed.

The coupling of a matter field to this gauge model raises intriguing 
questions. The repetition of the construction of the chiral scalar fields
outlined in the previous Sections indicates the possibility of constructing
chiral gauge fields. Suppose that a complex 
scalar field is introduced with higher derivative terms in the action. 
In the anti-ferromagnetic
phase of this model the helicity matter particle partners have different 
mass. In the low energy limit then one finds a chiral gauge theory 
for a single scalar chiral particle. Such a construction of a chiral
gauge theory by the dynamical breakdown of the space-time inversion
symmetry is a reminiscent of the Higgs mechanism where the spontaneous
breakdown of the internal symmetry is the key element in the construction
of the desired gauge theory.

The dynamics is fundamentally influenced by the anti-ferromagnetic gauge 
field even if the matter field does not support the
staggered order\cite{jochen}. One such a phenomenon is due to the one-loop effects
of the matter particle. 
The periodic gauge field vacuum creates forbidden zones in the
excitation spectrum in a manner analogous to the destructive Bragg 
reflection. Suppose that we have a massless fermion
whose Dirac sea level is placed in the middle of such a gap by the help of 
a suitable chosen chemical potential. The gap opening in the excitation spectrum
induces the dynamical breakdown of the chiral symmetry and leads to mass generation
without Higgs particles. Another interesting phenomenon is related to the
quantum fluctuations of the gauge field around the periodic vacuum. These
are the analogues of the phonons of the solid state lattice. When the
period length of the anti-ferromagnetic vacuum is renormalized to zero
then the continuum translation symmetry is not restored in the continuum limit.
So the phonons are not Goldstone modes and are not necessarily massless. 
When the length scale of the anti-ferromagnetic vacuum remains finite then
the massless phonons introduce an unscreened attractive interaction 
between the fermions which may lead to the formation of the Cooper pairs
and the superconductivity of the vacuum.

Finally we address the question whether the anti-ferromagnetic vacuum
what has been discussed so far in models with freely adjustable coupling constants
could actually be observed in realistic effective theories.
Suppose that we eliminate $n_f$ heavy fermions with mass $M$ coupled to a gauge field. 
The resulting contribution to the effective action what can be derived by the 
help of the hopping parameter expansion in lattice regularization 
is the sum of Wilson loops. The coefficient 
of a Wilson loop $\Gamma$ is 
$\beta_\Gamma=-n_f(-1)^{A(\Gamma)}O((aM)^{-L(\Gamma)})$,
where $A(\Gamma)$ and $L(\Gamma)$ stand for the area and the perimeter of the
Wilson loop in lattice units. The oscillating sign is the result of the fact that the 
product of the Dirac gamma matrices along a plaquette is -1. For an appropriate 
choice of $n_f$ and $M$ the semi-classical ground state can be made
anti-ferromagnetic by the help of the negative coefficient of the double 
plaquette term.

The above argument applies to the lattice models with heavy masses
but appears useless in continuum theories. The 
supercriticality\cite{qedsup} of the QED vacuum for ions with $Z>C_{cr}\approx173$ 
offers another non-perturbative mechanism without relying on the heavy
mass expansion in lattice regularization. In fact, for sufficiently large amplitude
oscillating electrostatic field the supercriticality occurs in each
potential valley periodically. Since the electron states are emptied at the
peaks of the potential and populated at the valleys the energy of the resulting
Dirac sea energy is lowered. Such a decrease of the fermion energy density
is in a good approximation independent of the period length of the vacuum. 
The oscillating Dirac see level "bends" the one particle wave functions
and increases the Dirac sea energy. But this increase is small if the
period length of the vacuum is long. The increase of the
electrostatic energy density is proportional to the square of the wave
vector of the vacuum. Thus for sufficiently strong and slowly varying
external periodic electrostatic field the vacuum should be non-homogeneous
\cite{nonhom}. 

It is amusing to realize that a slightly similar vacuum 
based on the bag model has already been proposed\cite{swva} for QCD.
The dynamical origin of this picture of the vacuum is the conjecture that 
the short range perturbative modes generate an effective theory
for the long range modes whose semi-classical vacuum is 
periodic. The non-perturbative aspects of the longe range dynamics are
supposed to take care of the homogeneity of the true vacuum, in a manner
similar to the spin liquids.

\section{Conclusion}
The goal of this talk was to show the serious impact certain higher order
derivative terms may have on the dynamics. If the scale of these higher order
terms remains finite during the renormalization and we have an effective theory
then phenomena, similar to Solid State Physics can be observed. If the
length scale of the higher order terms can be removed together with the
cut-off then qualitatively new continuum Quantum Field Theories are obtained
containing new type of particles by the help of a dynamical symmetry breaking
mechanism triggered by the kinetic energy. 

The phenomena outlined in this
talk are not yet well established, the renormalizability of the
scalar field theory model is known up to one-loop only. Substantial
work is left to achieve the level of accuracy which is usual for
ferromagnetic models. But we believe that this new anti-ferromagnetic phase 
has sufficiently interesting features which are warrant of presentation
even at this preliminary stage.
 
\section*{Acknowledgement}
I thank to Vincenzo Branchina, Jochen Fingberg and Herv\`e Mohrbach for the
collaborations and discussions which led to several results presented in 
this talk.

\end{document}